\documentclass[preprint, aps, showpacs]{revtex4-1}

\usepackage{graphicx}
\usepackage{amsmath,amssymb}
\usepackage[usenames,dvipsnames]{color}
\usepackage[colorlinks=true, citecolor=blue, linkcolor=WildStrawberry]{hyperref}

\newcommand{\drm}{{\rm d}}

\newcommand{\beq}{\begin{equation}}
\newcommand{\eeq}{\end{equation}}
\newcommand{\bdm}{\begin{displaymath}}
\newcommand{\edm}{\end{displaymath}}

\begin{document}

\title{Subtraction of Newtonian Noise Using Optimized Sensor Arrays}

\author{Jennifer C. Driggers}
\affiliation{LIGO Laboratory, Division of Physics, Math, and Astronomy, California Institute of Technology, 1200 East California Boulevard, MC 100-36, Pasadena, California 91125, USA}
\author{Jan Harms}
\affiliation{LIGO Laboratory, Division of Physics, Math, and Astronomy, California Institute of Technology, 1200 East California Boulevard, MC 100-36, Pasadena, California 91125, USA}
\author{Rana X Adhikari}
\affiliation{LIGO Laboratory, Division of Physics, Math, and Astronomy, California Institute of Technology, 1200 East California Boulevard, MC 100-36, Pasadena, California 91125, USA}

\begin{abstract}
Fluctuations in the local Newtonian gravitational field present a limit to high precision measurements, including searches for gravitational waves using laser interferometers. In this work, we present a model of this perturbing gravitational field and evaluate schemes to mitigate the effect by estimating and subtracting it from the interferometer data stream. Information about the Newtonian noise is obtained from simulated seismic data. The method is tested on causal as well as acausal implementations of noise subtraction. In both cases it is demonstrated that broadband mitigation factors close to 10 can be achieved removing Newtonian noise as a dominant noise contribution. The resulting improvement in the detector sensitivity will substantially enhance the detection rate of gravitational radiation from cosmological sources.
\end{abstract}

\maketitle


\section{Introduction}
\label{sec:intro}
Gravitational waves (GW) from astrophysical sources have the promise of revealing a rich new
vision of the universe~\cite{Sathya:LRR}. In the past decade, several kilometer sized 
terrestrial detectors of gravitational waves (such as TAMA300~\cite{TAMA:2008}, 
GEO600~\cite{LuEA2010}, Virgo~\cite{Virgo:2011}, and the Laser
Interferometer Gravitational-wave Obervatory (LIGO)~\cite{LSC2009b}), have
come online and made searches for signals in the 50-10000 Hz band. The reach of these
ground based detectors at low frequencies is limited by seismic and gravitational 
perturbations on the earth. A set of space missions (LISA~\cite{ESA:LISA}, 
DECIGO~\cite{Ando:DECIGO}) are being pursued to search for signals in the 10$^{-4}$--1 Hz band.

\begin{figure}[h]
  \centering
      \includegraphics[width=\columnwidth]{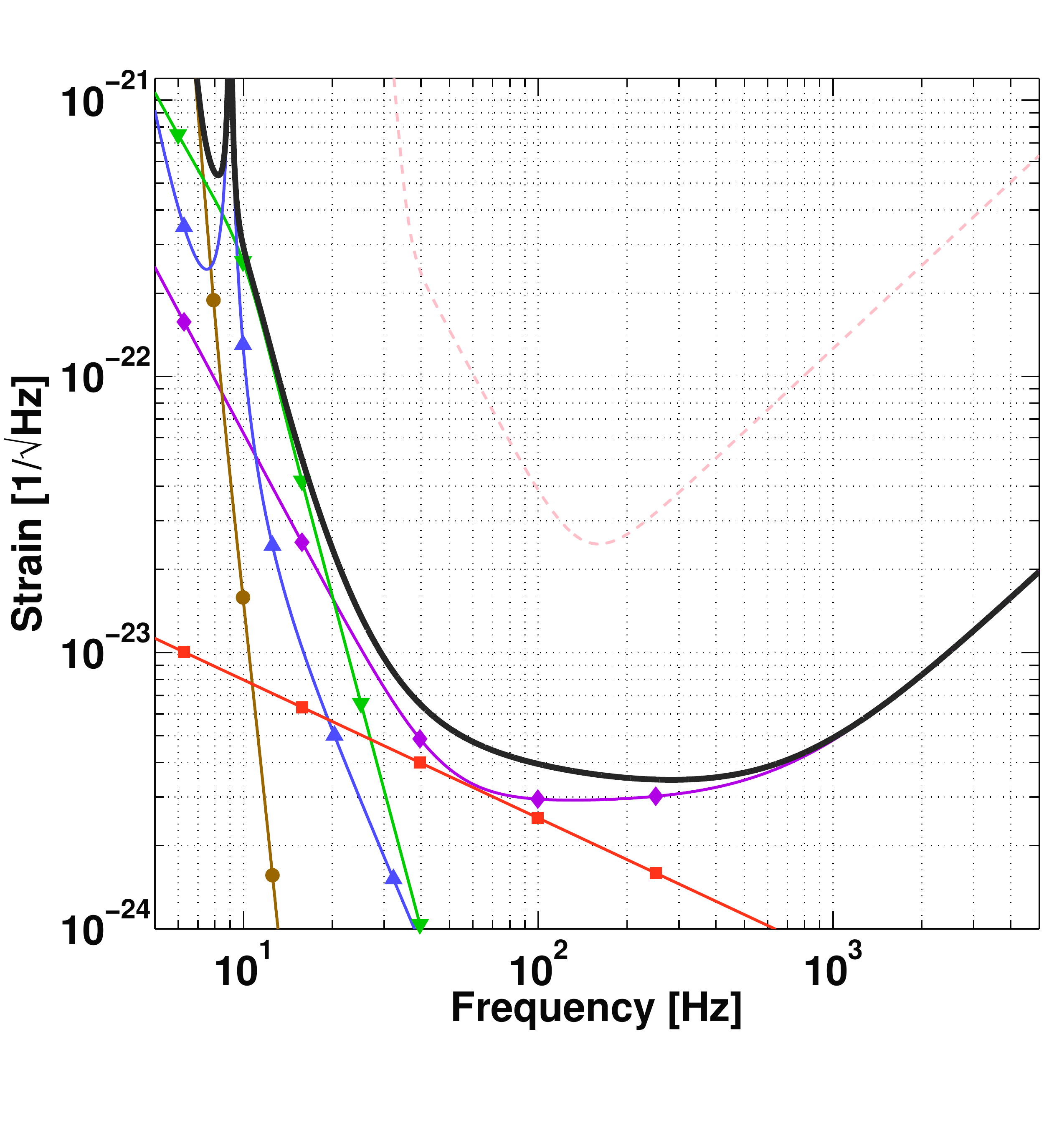}
      \caption{(color online). Strain noise spectral density of a second generation terrestrial detector—Advanced LIGO (black, bold line). The sensitivity of the first generation Initial LIGO (pink, dashed line) is shown for comparison. The Newtonian noise (green downward-pointing triangles) is dominating the Advanced LIGO sensitivity near 10 Hz. Other traces shown are other, nongravita- tional, limits to the sensitivity: direct seismic vibrations (brown circles), quantum radiation pressure and shot noise (purple dia- monds), mirror thermal noise (red squares), and mirror suspension thermal noise (blue upward-pointing triangles).}
    \label{fig:aLIGO}
\end{figure}

The multi-stage vibration isolation systems~\cite{aLIGO:SEI, Virgo:SAS} developed for
GW detectors should, in principle, be capable of reducing the direct influence of the ambient
seismic noise to below the quantum and thermodynamic limits of the interferometers. 
Unfortunately, there is no known way to shield the detectors' test masses from fluctuating
gravitational forces. As shown in Figure~\ref{fig:aLIGO}, our
calculations estimate that the fluctuations in the local Newtonian
gravitational field will be the dominant source of the mirror's
positional fluctuations near 10~Hz. This noise source has been referred to as gravity gradient noise or Newtonian noise (NN) in previous literature.

Early estimates of NN by Weiss~\cite{Rai:QPR}, Saulson~\cite{Sau1984}, and 
Hughes and Thorne~\cite{HuTh1998} have made increasingly better estimates of the
seismic environment and thereby, the Newtonian noise. In this work, we update estimates of Newtonian noise as well as describing a means to subtract its influence 
from the data stream.

\section{Newtonian noise budget for the LIGO sites}
\label{sec:appendix}
\begin{figure}[t]
  \centering
     \includegraphics[width=\columnwidth]{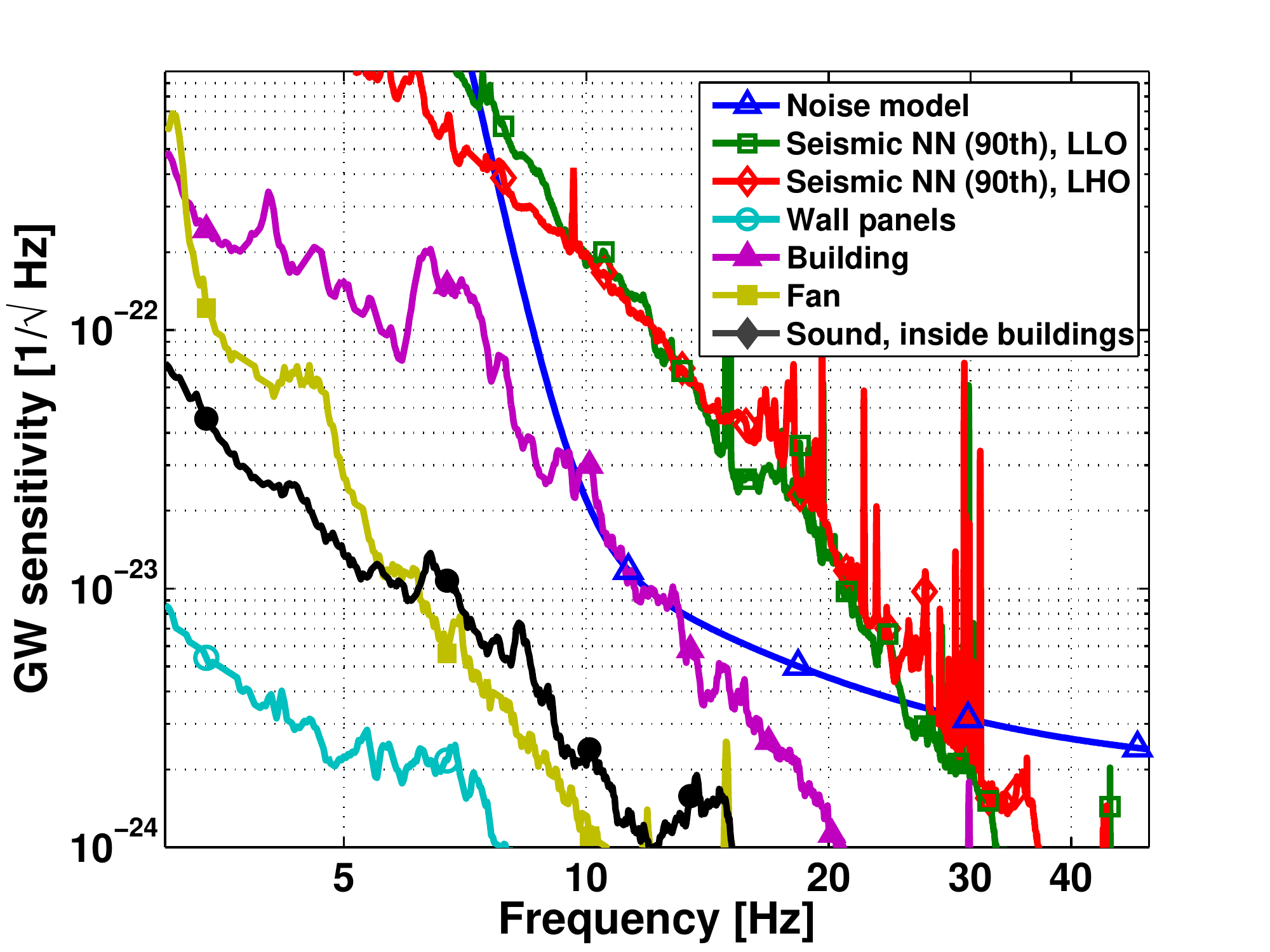}
\caption{(color online).    Seismic 90th percentile NN estimates for the LIGO Livingston (LLO) and Hanford (LHO) sites (green and red lines), third generation strain noise model (blue line), and additional NN estimates from vibrations of walls (cyan line), building tilts (magenta line), exhaust fans (beige line), and sound waves inside buildings (black line). Seismic contributions are the only NN source significant for third generation detectors and earlier. Building tilt will be important for detectors beyond the third generation, but is not a dominating noise source at this time.}
    \label{fig:nnbudget}
\end{figure}

In 2011, several measurements were carried out at the LIGO sites to define a Newtonian noise budget~\cite{DrigHarms:LLOnnMeas}. Accelerometers were used to monitor vibrations on water pipes, near exhaust fans, on top of the buildings and on the walls. Sound spectra were measured with microphones inside and outside of the LIGO buildings. The resulting NN estimates for each of these sources are summarized in Figure~\ref{fig:nnbudget}. In addition, the plot contains a representative noise model for potential upgrades to the advanced detectors such as Advanced LIGO and Advanced VIRGO, which we refer to as $3^{\rm rd}$ generation ground-based detectors~\cite{ISwhite2012}. 
Future detectors built at new sites, such as the proposed Einstein Telescope, we call $4^{\rm th}$ generation detectors~\cite{HiEA2011}. More specifically, the strain noise model (excluding NN) that we use to simulate interferometer noise for a 3$^{\rm rd}$ generation detector is:
\beq
\begin{split}
\sqrt{S_h(f)}=\bigg(&\frac{10^{-23}}{(f/10{\rm \,Hz})^{14}} \\
&+\frac{10^{-23}}{(f/10{\rm \,Hz})^2} +\frac{2\times10^{-24}}{(f/40{\rm \,Hz})^{1/50}}\bigg)\,\rm\frac{1}{\sqrt{Hz}}
\end{split}
\label{eq:noisemodel3G}
\eeq
It is based on current best estimates of technology advance to mitigate instrumental noise such as thermal noise, seismic noise, and quantum shot noise. The Newtonian noise is simulated as test-mass displacement noise. To convert from the displacement noise of a single mass into strain
noise, we multiply by 2 to account for the incoherent sum of 4 masses
and then divide by the interferometer length, $L = 4\,$km, to get strain. 

\begin{figure}[t]
  \centering
     \includegraphics[width=\columnwidth]{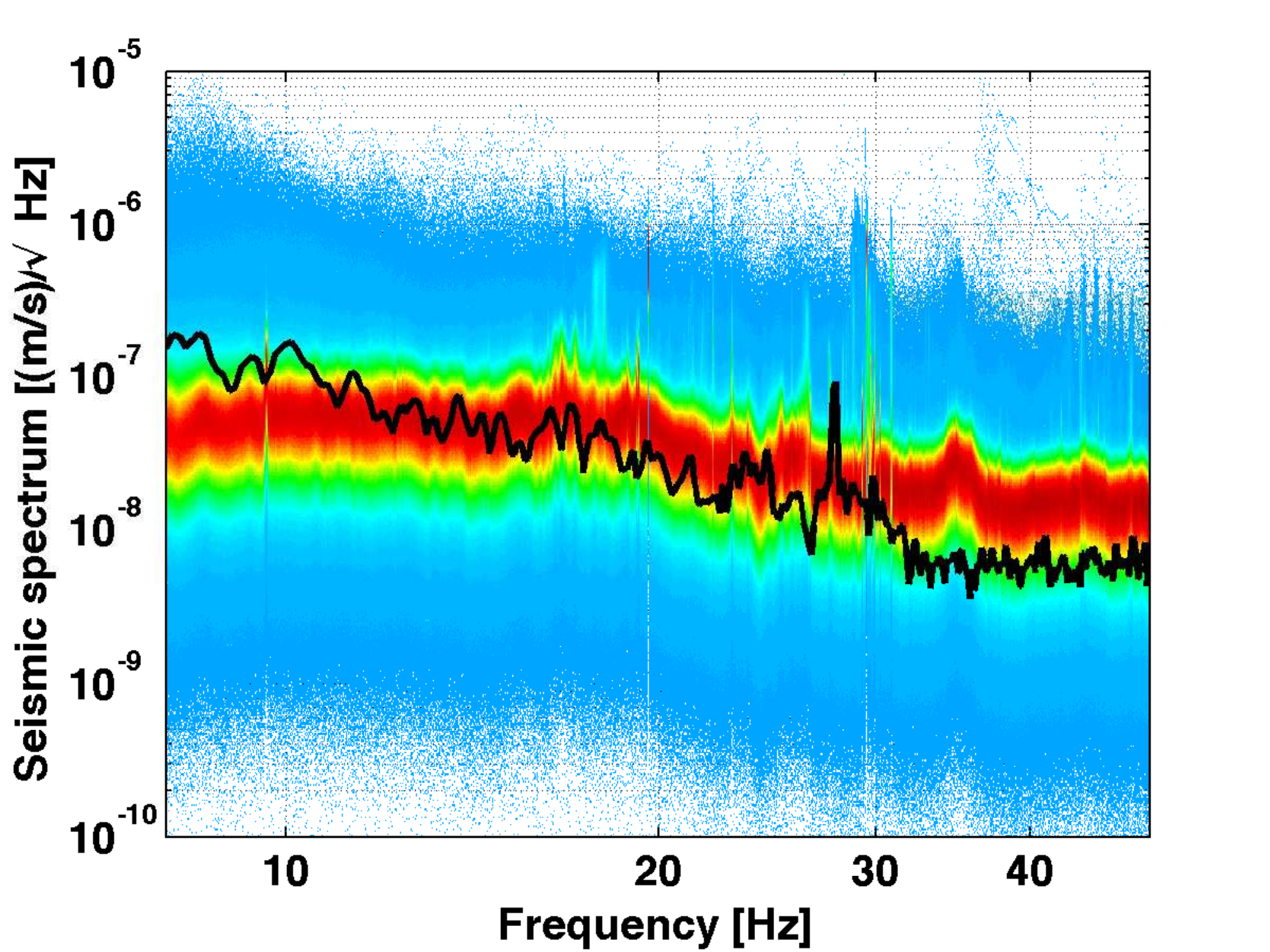}
    \caption{(color online). Histogram of one year of unaveraged 128\,s 
                   seismic spectra measured during the last LIGO science run 
                  inside the corner station of the Livingston detector. The black curve 
                  is the spectral density of the simulated seismic field. The spectral 
              histogram of the Hanford site is very similar for the frequencies plotted here.}
    \label{fig:seismicnoise}
\end{figure}

All measured curves in Figure~\ref{fig:nnbudget} are derived from 90$^{\rm th}$ percentiles of spectral histograms similar to the one shown in Figure~\ref{fig:seismicnoise} for the seismic measurement at the LIGO Livingston site. Note that the seismic curves for both LIGO sites presented in Figure~\ref{fig:nnbudget} are more recent, using a more accurate, non-averaged, analysis of the seismic percentiles, as compared to~\cite{DrigHarms:LLOnnMeas}.

According to these estimates, seismic NN is the only significant Newtonian noise contribution for 3$^{\rm rd}$ generation and earlier detectors, so other contributions to NN such as 
building vibrations or air pressure fluctuations are not considered in the following.


\section{Simulation of Seismic Newtonian Noise}
\label{sec:simulation}
Since Newtonian noise cannot, at this time, be directly measured, 
we must base our estimates of subtraction capabilities on simulated noise.
We attempt to obtain a sufficiently
 accurate estimate of the NN based on information about its source,
 which, in this case, is the seismic field. 

In this section, we give a description of a time-series generator for seismic fields, the associated NN, and other instrumental noise of the interferometer and seismometers. We also discuss the suitability of our simulation as an estimate of NN at the LIGO sites.  The problem is set up as a full time-domain simulation of seismic fields and instrumental noise. Instrumental noise such as seismometer noise or test-mass displacement not generated by NN is treated as stationary. In contrast, we do not assume stationarity of the seismic field. Here the attempt is to simulate fields with seismic spectra that are comparable to seismic spectra measured at the sites, and to make the field composition as complex as possible in order to test NN subtraction schemes on challenging scenarios. Still, due to computational limitations, simplifications are necessary. In the general case, if ground displacement $\vec \xi(\vec{r}, t)$ is weak, then the test-mass acceleration due to NN can be estimated by the integral over the entire ground medium
\beq
\delta \vec a_{\rm NN}(\vec{r}_0, t) = G \int \drm V\, \frac{\rho_0(\vec{r})}{|\vec{r} - \vec{r}_0|^3} \left(\vec\xi (\vec{r}, t)-3(\vec{e}_r\cdot\vec\xi (\vec{r}, t) ) \vec{e}_r\right)
\label{eq:aNNintegral}
\eeq
where $\rho_0$ is the density of the ground, $G$ is the gravitational constant, $\vec{r}_0$ the position of the test mass, $\vec{r}$ points to locations in the ground, and $\vec{e}_r$ is the unit vector pointing from $\vec{r}_0$ to $\vec{r}$~\cite{HaEA2009a}. This equation is valid for arbitrary seismic fields and represents the noise imprinted on the test mass due to NN. In our simulation, we only consider seismic fields composed of surface waves. This simplification is enforced by computational limitations since generating NN time series from simulated 3D seismic fields would require months-long simulation runs. We expect this assumption to be reasonable, since surface waves are expected to have much larger amplitudes than body waves~\cite{HuTh1998}, and so surface waves give the dominant contribution to NN at the surface, however seismic array measurements currently in progress at the LIGO sites will put this assumption to the test.

Freely propagating surface waves like Rayleigh waves and their overtones produce NN in such a way that there is always an effective 2D representation of the problem (which is not generally true for all supported wave fields, such as scattered waves).  This implies that the numerical simulation can be set up as a 2D simulation, which is why Equations~\ref{eq:wavelet} and~\ref{eq:local} only describe vertical displacement.  This approach was chosen to reduce computational costs, and does not over simplify the subtraction problem as long as scattering of seismic waves is weak.

The simulated seismic field is composed of two wave types; wavelets and symmetric surface waves.  Wavelets represent seismic waves from far-field sources, while symmetric surface waves represent disturbances due to local sources. The vertical displacement due to a wavelet is described by
\beq
\xi(\vec r,t) = \xi_0\exp(-\tau^2/(2 \Delta T)^2)\cos(2\pi f\tau+\phi_0)
\label{eq:wavelet}
\eeq
with $\tau = t-\vec k\cdot(\vec r-\vec r_0)/(2\pi f)$. Twenty wavelets are injected for each second of time series randomly distributed over the entire simulation time so that wavelet numbers can be different each simulated second. Frequencies $f$ are drawn from a uniform logarithmic distribution between 8\,Hz and 30\,Hz, which includes the full range of frequencies for which NN is expected to be dominant. Wavelet durations $\Delta T$ are uniformly distributed between $10/f$ to $20/f$, to represent that wavelets can vary in duration depending on the type and source of the disturbance. The distribution of wave vectors $\vec k$ is isotropic, to represent far-field sources from all directions. The average speed of sound for seismic waves in the ground is approximately 200\,m/s~\cite{HaOR2011} so we allow seismic speeds in our simulation to vary uniformly from 195\,m/s to 205\,m/s. This variance in speed is a brute-force method to simulate wave scattering, but it is very likely an overestimation of the effect. The initial location $\vec r_0$ of the wave maximum lies in the direction of the back-azimuth of the incident wave such that the wave is guaranteed to reach the location of the test mass within the simulated duration of the time series. The initial phase $\phi_0$ of any single seismic wave is not a critical parameter.  What is important is that not all seismic fields in our simulation have the same phase, so $\phi_0$ is drawn from a uniform distribution between 0 and $2\pi$. 

The second type of wave in the simulation is the symmetric surface wave, described by Bessel functions, with vertical displacement
\beq
\xi(\vec r,t) = \xi_0 J_0(k_0 R)\cos(2\pi ft+\phi_0)
\label{eq:local}
\eeq
with $R = |\vec r-\vec r_0|$. Equation~\ref{eq:local} represents fields from sources located at $\vec r_0$ with distance $r_0$ drawn uniformly between 10\,m and 20\,m.  Sources more distant than 20\,m appear at the test mass as distant sources, represented by wavelets as in Equation~\ref{eq:wavelet}. All other parameters are obtained in the same way as for the wavelet, where as before the variation in seismic speed leads to a corresponding variation of the wave number $k_0$. We assume that local sources do not vary strongly over the relevant time scales (defined by the subtraction procedure; see following Sections), so that the local sources are considered stationary. A fixed number of 10 waves from local sources is used.  Below 1\,Hz, there are typically no more than 2 waves present at a time~\cite{ToLa1968}, so we expect that, while the number of waves present can increase with the frequency of the seismic waves, twenty distant and ten local sources is a conservative overestimate of the complexity of seismic fields we will see at the LIGO sites.  Seismic array measurements currently underway at the LIGO sites are expected to confirm this assumption.

The full simulation covers a surface area of 100\,m $\times$ 100\,m with the test mass at its center, which is larger than the area from which interesting NN contributions are expected~\cite{HaEA2009a}. The number of grid points along each dimension is $N=201$ so that the grid-point spacing is 0.5\,m. We choose a $201\,\times\,201$ point grid as a compromise between overall grid area, grid spacing, and computational time.  The test mass is suspended 1.5\,m above ground, which is approximately the height of the LIGO test masses.  As the effective 2D representation is based on the surface term of the gravity perturbations and not the full dipole form~\cite{HaEA2009a}, we convert the integral in Equation~\ref{eq:aNNintegral} into a discrete sum over grid nodes.  Using only the surface contribution to the integral, the test mass acceleration along the direction of the interferometer arm is
\beq
a_{\rm NN}^{\rm arm}(t) = G\rho_0 \drm S \sum\limits_{l=1}^{N\times N}\frac{\xi_l(t)}{r_l^2}\cos(\theta_l)
\label{eq:aNN}
\eeq
where $\rm dS$ is the area of the square enclosed by four neighboring grid points, $\xi_l(t)$ is the vertical displacement of grid point $l$ at time $t$, $r_l$ is the distance between the grid point and the test mass, and $\theta_l$ is the angle between the vector pointing from the test mass to the grid point and the direction of the interferometer arm. The sum over grid points in Equation~\ref{eq:aNN} is used to determine the time series of the NN at the test mass.  Time series for each seismometer in Sections~\ref{sec:postsub} and~\ref{sec:feedforward} are calculated separately using Equations~\ref{eq:wavelet} and~\ref{eq:local}, so seismometer locations are not restricted to coinciding with grid points.

We utilize models of the instrumental noise of seismometers and the strain noise of an interferometer to more accurately determine the NN subtraction efficiency, as described in Sections \ref{sec:postsub} and \ref{sec:feedforward}.  The instrumental noise of all seismometers is simulated with spectral densities that are white (frequency independent) in units of velocity and have a value of $10^{-10}\,\rm m/\sqrt{Hz}$ at 10\,Hz. This is a
conservative estimate for commercial geophones.

The seismic spectrum itself plays a minor role for the purpose of this paper, but 
nevertheless we defined distributions for $\xi_0$ in Equations~\ref{eq:wavelet}
and~\ref{eq:local} in such a way that the spectral density approximates the median 
spectrum measured at the LIGO sites. The plot in Figure~\ref{fig:seismicnoise} shows 
the histogram of unaveraged 128\,s spectra measured at the LIGO Livingston site 
over a time of one year during the last science run, and the black curve represents 
the average spectrum of the simulated seismic field.

The average spectral density derived from the histogram is about a factor of 2 to 3 larger at frequencies between 10\,Hz and 30\,Hz than the model used in~\cite{HuTh1998}, with a correspondingly larger NN spectrum. 

The sampling frequency for all time series is $f_{\rm s}=100\,$Hz and the observation time is $T=100\,$s. We plan to test our subtraction techniques on longer duration simulated data in the future, however computational time restraints have kept us to this moderate duration for the time being.  All time series are high-pass filtered with corner frequency $5\,$Hz directly after being generated to avoid numerical problems.
\begin{figure}[t]
  \centering
     \includegraphics[width=\columnwidth]{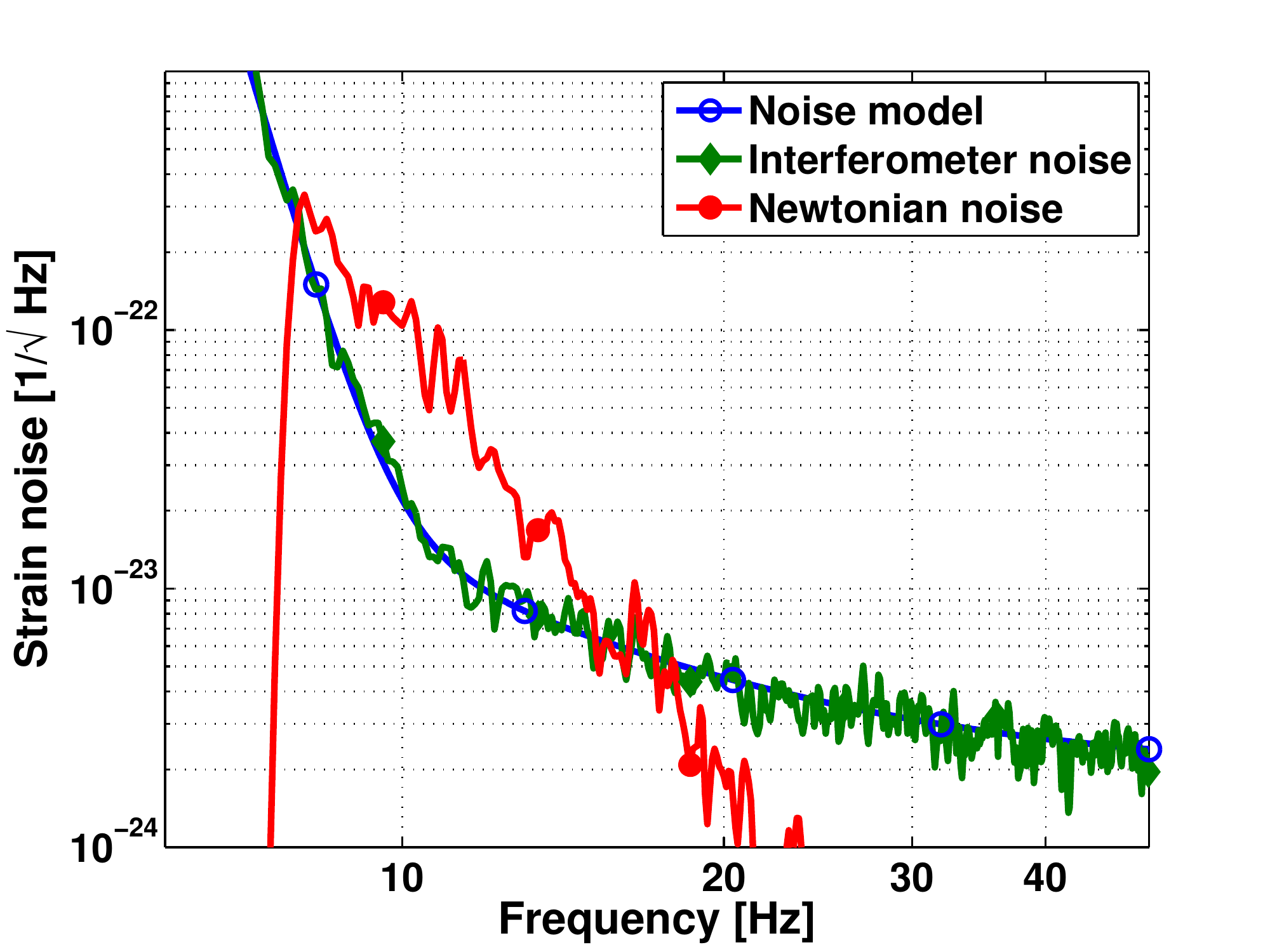}
    \caption{(color online).    Spectrum of simulated Newtonian noise (red line), reference third generation sensitivity curve described in Eq. (1) (blue line), and simulated interferometer noise based on this noise model (green line). Since other noise sources such as mirror suspension thermal noise and direct seismic vibra- tions will be the limiting noise sources for second and third generation detectors below $\sim8\,$Hz, we do not need to consider NN at these low frequencies. We therefore do not include low frequency information in our NN estimate, which creates a sharp cutoff when the simulated data is viewed in the frequency domain}
    \label{fig:samplespectrum}
\end{figure}
As can be seen in Figure~\ref{fig:samplespectrum}, interferometer noise dominates NN below 8\,Hz and above $\sim$20\,Hz so that we can safely ignore frequencies outside this range when testing NN subtraction methods. 


\section{Sensor Array Optimization}
\label{sec:optimize}
Optimization of seismic arrays with respect to NN subtraction was discussed in~\cite{BeEA2010}. The authors calculated subtraction residuals analytically by evaluating explicitly the correlation between seismometers and the test mass as a function of seismometer locations. The average subtraction residual can be written as
\beq
R=1-\frac{\vec C_{\rm SN}^\mathrm{\,T}\cdot (C_{\rm SS})^{-1}\cdot\vec C_{\rm SN}}{C_{\rm NN}}
\label{eq:residual}
\eeq
Here, $\vec C_{\rm SN}$ is the cross-correlation vector between seismometers and the NN acceleration of the test mass, $C_{\rm SS}$ is the cross-correlation matrix between seismometers, and $C_{\rm NN}$ is the NN variance. These quantities can also be interpreted as (cross-)correlation spectral densities. Given a fixed number of seismometers, the optimal array is found by changing seismometer locations and minimizing $\sqrt R$. The equation is idealized as it does not depend on any details about the way subtraction is implemented,
i.~e.~whether a finite impulse response (FIR) filter is used, or some non-causal post-subtraction filter (see following two sections for details). For this reason it describes the performance of all subtraction methods that are based on linear filtering, and the optimal array found by minimizing $\sqrt R$ is universal for all linear noise filters. Since it is very likely that different noise cancellation techniques will be combined in practice, it seems that optimization based on Equation~\ref{eq:residual} is the best one can do.

\begin{figure}[t]
\centering
\includegraphics[width=\columnwidth]{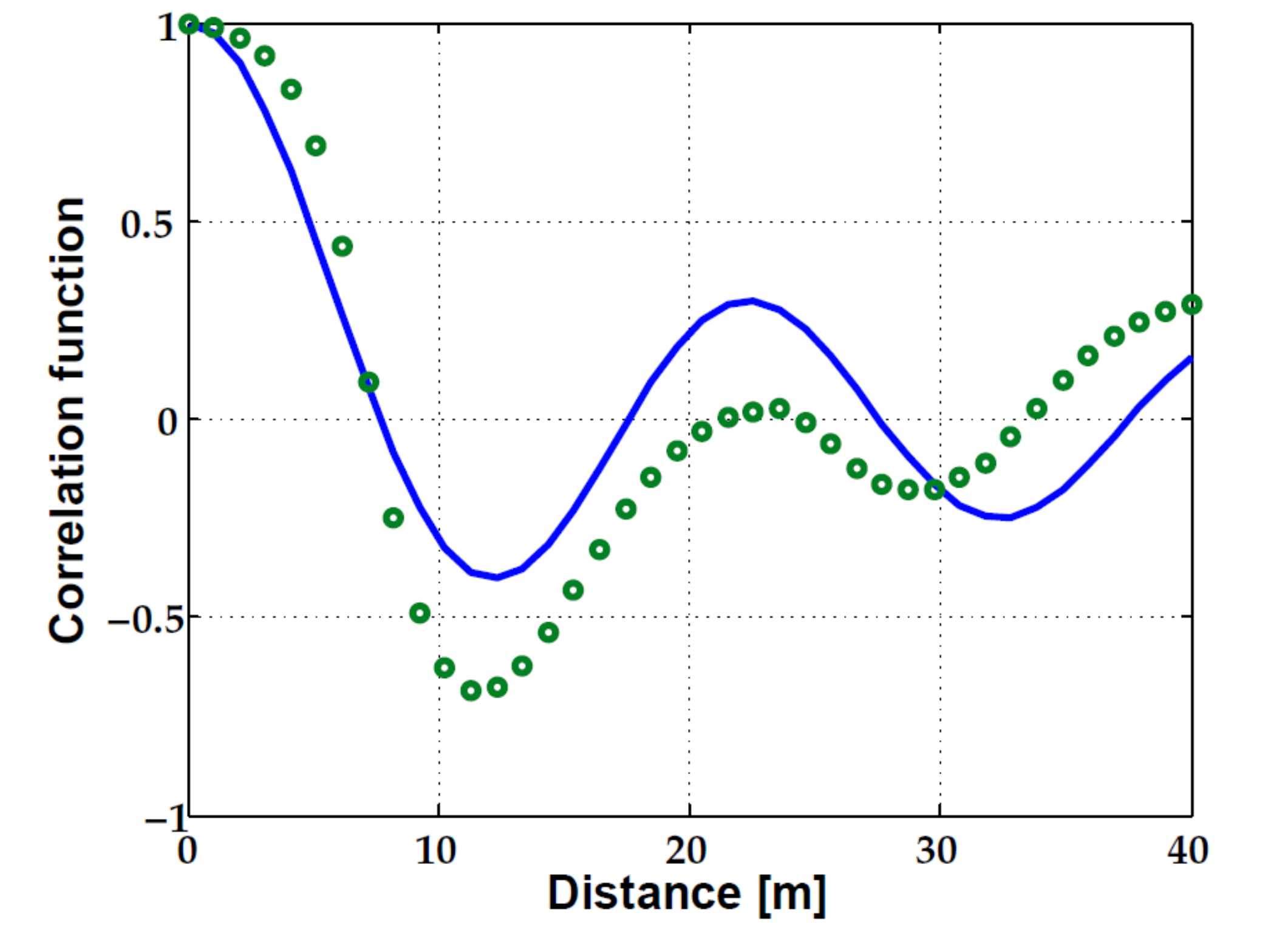}
\caption{(color online). Comparison between the theoretical model for seismic correlation of isotropic plane-wave surface fields as described by Equation~\ref{eq:CSS} (solid line), and the correlation calculated by the simulation of a field composed of wavelets and locally generated waves (dotted line). Contributions from local sources cause variations of the correlation curve at larger distances between simulation runs (see Section~\ref{sec:simulation} for details). Therefore, theoretical and simulated correlations match at small distances and deviate more strongly at larger distances. Since seismic fields in the context of NN subtraction only matter very near the test mass, the match between simulated and theoretical correlations at small distances means that the optimal array determined analytically by minimizing $\sqrt R$ in Equation~\ref{eq:residual} should also perform well in simulation, and more importantly that the simulation should be representative of our real subtraction ability.}
\label{fig:comparebessel}
\end{figure}

Correlation patterns of surface waves observed in nature are often well approximated by Bessel functions that characterize isotropic plane-wave surface fields~\cite{PLB2009,StEA2009}. Adopting a more convenient normalization the corresponding seismic correlation $C_{\rm SS}$ between two points $\vec r_i,\,\vec r_j$ on the surface is given by
\beq
C_{\rm SS}(\vec r_i,\vec r_j)=J_0(2\pi |\vec r_i-\vec r_j|/\lambda)+\frac{1}{{\rm SNR}^2}\,\delta_{ij}
\label{eq:CSS}
\eeq
where $\lambda$ is the length of the seismic wave, and SNR is the signal-to-noise ratio of the seismometers. To find out how well this theoretical model approximates the seismic correlation in the simulation, we calculated $C_{\rm SS}$ between seismometers of increasing distance using our simulated seismic fields. The result is shown in Figure~\ref{fig:comparebessel}, where we show that the correlation vs. distance of our simulated seismic fields match the theoretical correlation of seismic fields fairly well, albeit not precisely.

Other terms in Equation~\ref{eq:residual}, using the same normalization as for Equation~\ref{eq:CSS}, are the NN variance 
\beq
C_{\rm NN}=0.5
\eeq
and the correlation between seismic displacement and NN acceleration of the test mass located at the origin
\beq
C_{\rm SN}(\vec r_i)=J_1(2\pi r_i/\lambda) \frac{x_i}{r_i}
\label{eq:CSN}
\eeq
with $r_i=|\vec r_i|$, and $x_i$ is the projection of $\vec r_i$ onto the direction of the interferometer arm. Since Equation~\ref{eq:residual} is independent of seismic or NN amplitudes, we can use any suitable normalization of the seismic field or the NN.

Finding the optimal array is not a trivial task. The result of a stepwise optimization by placing one seismometer after another leads to array configurations very different from the optimum. For the model described by Equations~\ref{eq:CSS} to~\ref{eq:CSN}, the stepwise optimization yields a straight line of seismometers along the direction of the arm, approximately symmetric about the test mass, independent of the number of seismometers. 
\begin{figure}[t]
\centering
\includegraphics[width=\columnwidth]{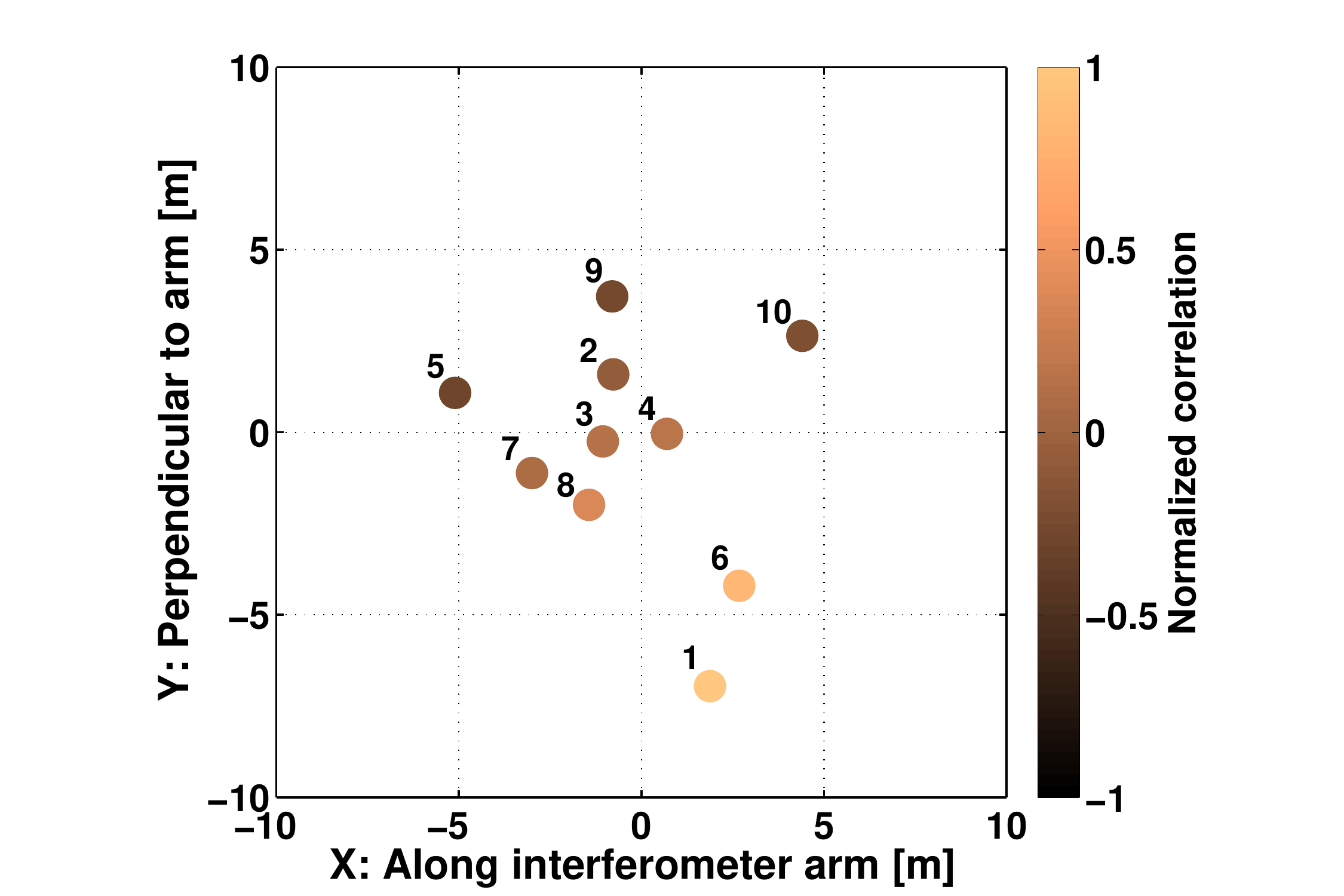}
\caption{(color online). Locations of 10 sensors resulting from numerically minimizing the subtraction residual. The optimal array should be symmetric about the test mass located at (0,0), but the subtraction residual is less than 10$^{-6}$ at 10\,Hz for this array. The colors indicate the normalized seismic correlation between seismometer 1 and all other seismometers.}
\label{fig:OptimumArray}
\end{figure}
Therefore, configurations close to the optimum can only be found by optimizing all seismometer locations simultaneously. A systematic numerical search for the optimum for more than a few seismometers is prohibitively computationally expensive, and approximate numerical optimization methods need to be applied. The array configuration that we call optimal in the following sections is shown in Figure~\ref{fig:OptimumArray}. It was found numerically by running a particle-swarm minimization code~\cite{PSOarticle, PSOcode} to optimize the location of 10 noiseless seismometers. It should be clear that the optimal array should have some kind of symmetry, so we know that this configuration is suboptimal. The optimization was stopped at a residual $\sqrt{R} \sim 10^{-6}$ at 10\,Hz, which is negligible for all practical purposes. So while this configuration does not represent a global optimum, its subtraction performance should be sufficient for Advanced LIGO and 3$^{\rm rd}$ generation detectors. As many configurations yield similarly small subtraction residuals, we added further components to the cost function $\sqrt{R}$ to make sure that seismometers are not placed too close to each other.  The array shown in Figure~\ref{fig:OptimumArray} is the result of minimizing this combined cost function.

\begin{figure}[t]
\centering
\includegraphics[width=\columnwidth]{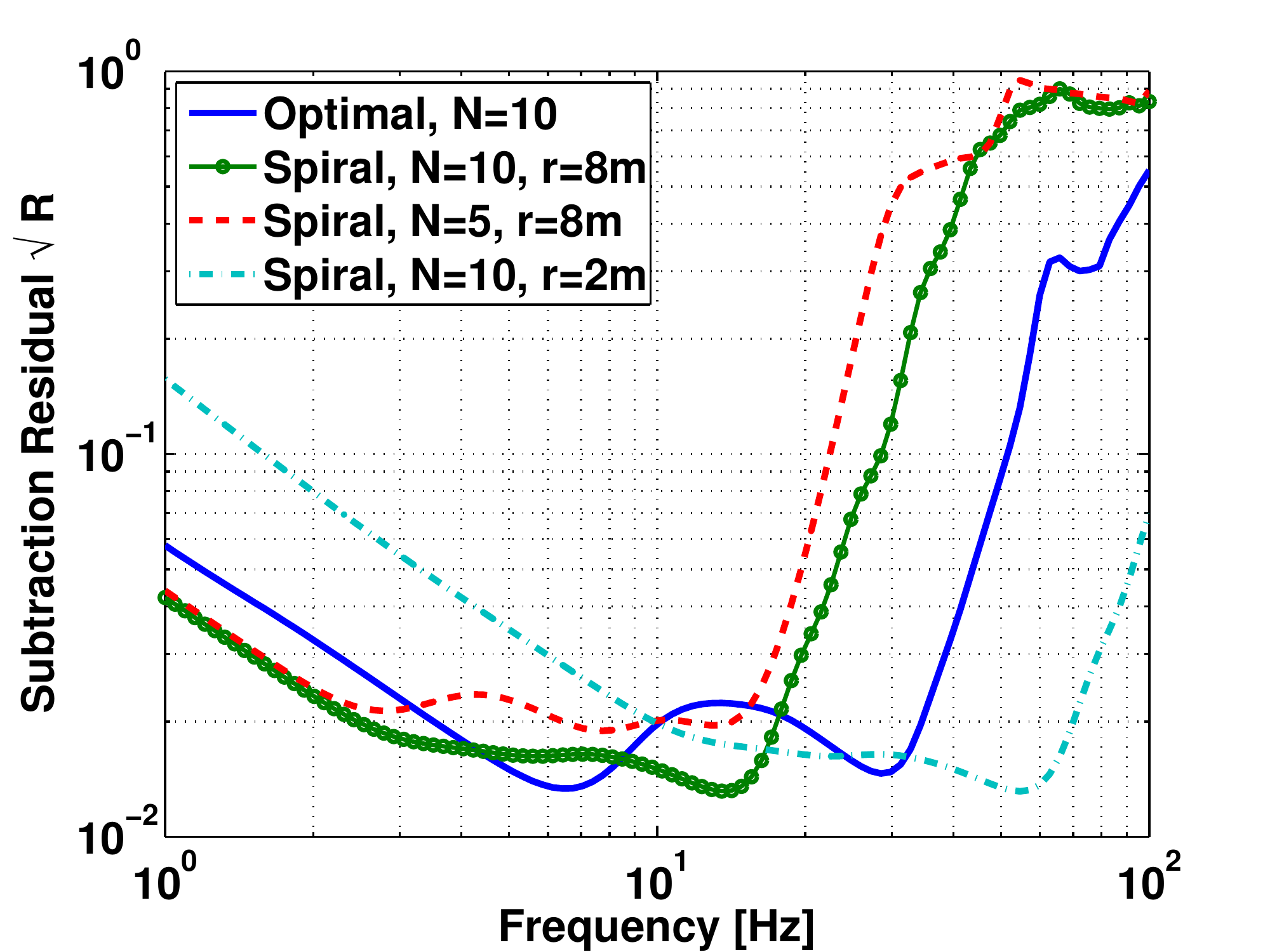}
\caption{(color online). Subtraction residual as defined in Equation \ref{eq:residual} vs. frequency for the array shown in Figure~\ref{fig:OptimumArray}, and three different spiral configurations. The `N=10, r=8\,m' array is shown in Figure~\ref{fig:bode}, and the `N=10, r=2\,m' array is the same, but with all seismometer coordinates scaled down by a factor of 4.  The `N=5, r=8\,m' array has two sensors at the same positions as numbers 1 and 10 in Figure~\ref{fig:bode}, and 3 other sensors distributed along the two-turn spiral in between these two. It is assumed that the seismometers measure ground motion with $\rm SNR=100$ at all frequencies. The Rayleigh-wave speed is 200\,m/s.}
\label{fig:OptimumEpsilons}
\end{figure}

As one can see from Equations~\ref{eq:CSS} to~\ref{eq:CSN}, the residual $R$ is a function of seismic wavelength, and therefore frequency, and broadband subtraction performance needs to be investigated. The subtraction residual of the array in Figure~\ref{fig:OptimumArray} was minimized at 10\,Hz for a seismic wave speed of $200$\,m/s. In Figure~\ref{fig:OptimumEpsilons} we show the subtraction residual as a function of frequency for various array configurations.
One can see how the number of seismometers and the array size affect subtraction residuals. It is clear that a very small array does not perform well at low frequencies since it provides highly degenerate information at these frequencies whereas larger arrays sample a larger part of the seismic wave. A smaller number of seismometers simply leads to a broadband increase of subtraction residuals except for the smallest frequencies. We want to emphasize that these theoretical predictions only hold approximately for the numerical simulation presented in the following sections, since it does not account for details of the subtraction method, as discussed previously.
  
Note that all arrays discussed here refer to sensors placed on the ground inside the LIGO buildings.  In-chamber vibrations are already suppressed to the level of the noise floor of the best available sensors, so we are not able to measure any motion relevant to NN inside the LIGO vacuum envelope.  Thus, we use ground-mounted sensors to measure NN outside the vacuum envelope.


\section{Offline Post-Subtraction}
\label{sec:postsub}
For the purpose of this paper, offline post-subtraction denotes the cancellation of noise in recorded data. The noise cancellation filter can therefore be causal or acausal. In this section, we will present a simple acausal implementation of the post-subtraction. The method chosen here is to cancel NN on short segments of recorded interferometer time series. The basic idea is to optimally construct a vector of filter coefficients, one coefficient per seismometer, and then use these coefficients to form a linear superposition of the seismometer channels as the NN estimate.  For a general introduction to digital filtering techniques, please see, for example,~\cite{Pri2001}.  In order to determine the effectiveness of this offline subtraction, we look at the residual interferometer sensitivity after removing the NN:
\beq
D(t_m,t_{m+1}) = I(t_m,t_{m+1})-\frac{\langle I^{\rm c},\vec S^{\rm c}\rangle_m}{\langle \vec S^{\rm c},\vec S^{\rm c}\rangle_m}\cdot \vec S^{\rm c}(t_m,t_{m+1})
\label{eq:postsub}
\eeq
The residual $D$ in Equation~\ref{eq:postsub} corresponds to the interferometer data $I$ minus the NN estimate from the seismometer data. The time series used to calculate the NN estimate are pre-conditioned with whitening and band-pass filters focussing on the 8\,Hz to 30\,Hz NN band before the correlations are evaluated. We also found it necessary to apply an anti-aliasing window (we used the high gain Nuttall window) for reasons that will be described below. All quantities subject to the pre-conditioning are marked with a ``$c$''. $\langle I^{\rm c},\vec S^{\rm c}\rangle_m$ denotes the vector of cross correlations between the interferometer data $I^{\rm c}$ and all seismometers $\vec S^{\rm c}$ using data of segment $m$ acquired between $t_m$ and $t_{m+1}$. Similarly, $\langle \vec S^{\rm c},\vec S^{\rm c}\rangle_m$ is the cross-correlation matrix between all seismometers. This means that the filter used here will have one filter coefficient per seismometer for the entire time interval $t_m$ to $t_{m+1}$. 

We must determine a reasonable time duration for each segment. Segments are too short if the spectral resolution is too small to disentangle seismic waves at different frequencies.  Segments may be too long if the number of seismic waves in that time frame becomes large.  A Wiener filter that sees many seismic waves may begin to average over the different waveforms and provide non-optimal noise suppression.  Choosing the goldilocks segment duration is somewhat arbitrary, however it is likely that the appropriate duration depends as much on the nature of the seismic field as on the frequency band targeted by the filter. With our simulation we found the best subtraction performance for 2\,s long segments.  This is an acausal technique, so testing can be done offline to determine the duration for which we see maximal NN suppression on the real data.

Since filter coefficients are re-evaluated for each segment $m$, a simple subtraction of NN estimates from consecutive segments can lead to discontinuities in the residual time series. For this reason the Nuttall anti-aliasing window is applied so that noise subtraction is suppressed at the beginning and end of a time segment. Consequently time segments are defined with overlap to provide continuous subtraction of NN. Using the Nuttall window, we found excellent subtraction performance with 0.3 fractional segment overlap. Again, some investigation can be done to optimize this number for real data in the future.

\begin{figure}[t]
\centering
\includegraphics[width=\columnwidth]{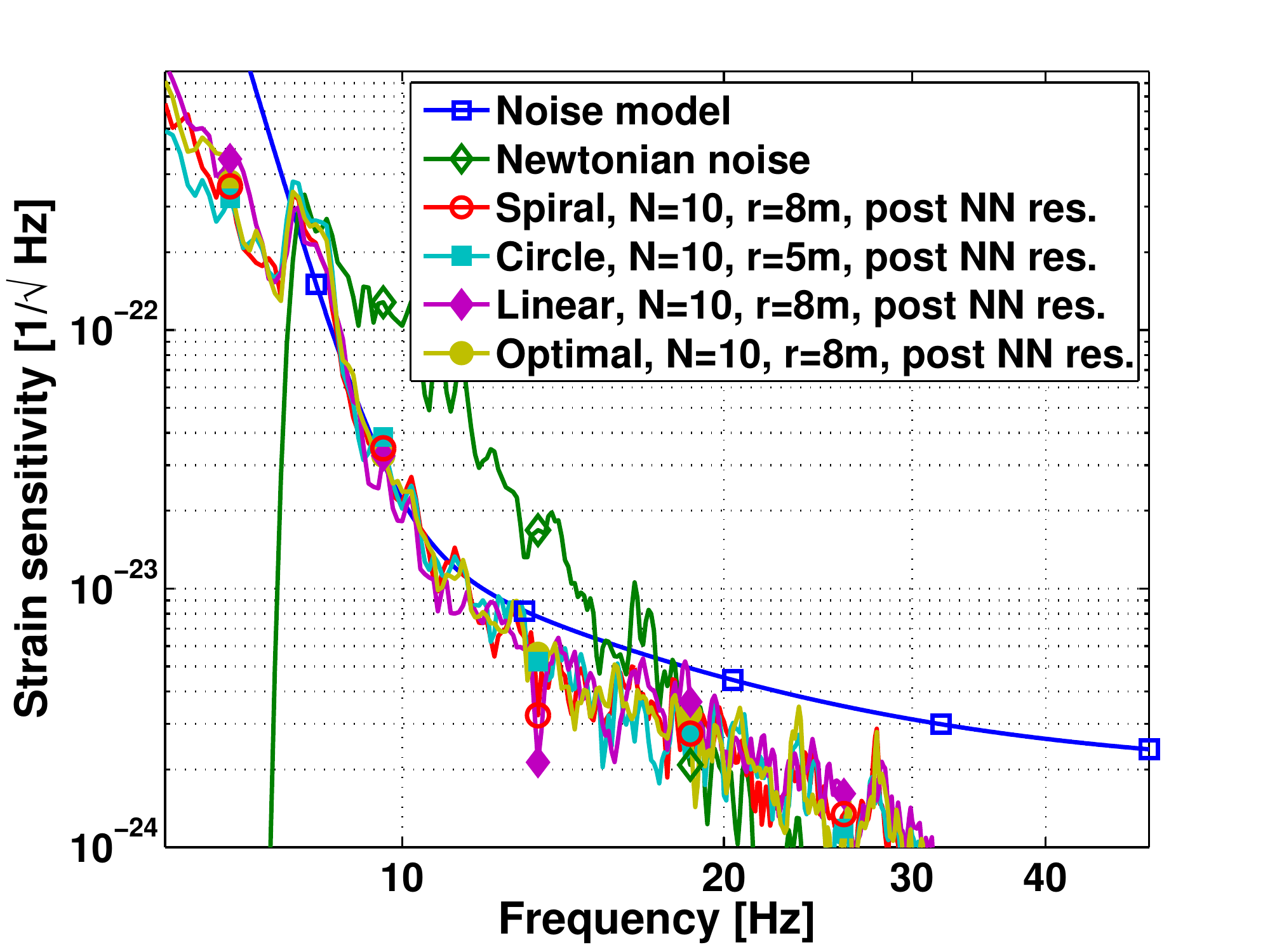}
\caption{(color online). Offline Newtonian noise subtraction efficiency for third generation detectors. Spectrum of simulated Newtonian noise (green line), proposed third generation sensitivity curve (blue line), and NN residuals of postsubtraction for a spiral array (red line), circular array (cyan line), linear array (magenta), and the optimal array (beige line). Filters derived from all four arrays reduce the simulated NN to a level below other sources of interferometer noise as represented by the noise model.}
\label{fig:postsubtraction}
\end{figure}

Optimal array design has already been discussed in Section~\ref{sec:optimize}. We will compare the subtraction performance of the optimal array presented there with a circular, a spiral, and a linear array. All arrays contain 10 seismometers, and are optimized in terms of the extent of the array relative to the location of the test mass. The linear array is simply a line of uniformly spaced seismometers along the direction of the arm extending 8\,m away from the test mass in both directions. This linear array is slightly different from the result of the stepwise optimization discussed in Section~\ref{sec:optimize}, but the subtraction residuals are similar.  The circular array consists of one seismometer under the test mass and 9 seismometers in a circle of radius 5\,m around the test mass. The configuration of the spiral array is shown in Figure~\ref{fig:bode} of the following Section. The residuals of the noise subtraction (described in Equation~\ref{eq:postsub}) for each array are shown in Figure~\ref{fig:postsubtraction}. 
The noise model represents the sensitivity curve of a potential upgrade of the advanced detectors not including the NN, as described by Equation~\ref{eq:noisemodel3G}. Approximately, all arrays perform equally well in post-subtraction. The goal to reduce the NN residuals to a level below the noise model is achieved over the entire NN band except for the very smallest frequencies. In Section~\ref{sec:feedforward}, we will investigate the possibility of combining the post-subtraction with an online feed-forward cancellation.


\section{Online Feed-Forward Subtraction}
\label{sec:feedforward}
Online feed-forward (FF) subtraction can be implemented in two ways. It is possible to continuously cancel NN by exerting a cancellation force directly on the test masses. Alternatively, the cancellation can be done on interferometer data. If we had ideal, noise-free actuators, the residuals resulting from applying forces on the test masses and online FF cancellation applied to the data would be the same.  Applying hardware cancellation forces could also be used to suppress the problem of any non-linear response of the detector to strong NN forces, but is very technically challenging to implement~\cite{FFW}. Since we do not believe that near-future detectors will suffer from significant non-linear upconversion due to NN, we only consider FF cancellation applied to the data. 

The main difference between online FF and post-subtraction is that online subtraction can only be done with causal filters. Furthermore, the FF filter coefficients can only change slowly in time following slow changes of average correlations between seismometers and the test mass.

The FF subtraction scheme that we propose is based on a multi-input, single-output (MISO) finite impulse response (FIR) filter that is continuously applied to the interferometer output to filter out the NN as was already demonstrated successfully for seismic noise cancellation schemes~\cite{Drig2012wiener}. The inputs consist of the seismometer channels, and the single filter output is the NN estimate. 

Average correlation between seismometers and interferometer data has a predictable form since average properties of the seismic field depend solely on the wave composition of the seismic field, which is characteristic for each site. This correlation pattern was investigated in Section~\ref{sec:optimize}, where we showed that the simplest theoretical model is a good representation even for the more complex wave composition that is used in our numerical simulation. 

As we will show in the following, sufficient FF subtraction down to the level of other noise contributions can be achieved with a variety of array configurations including arrays that have seismometers with negligible correlation with the test mass NN. The more important design factors are the number of seismometers and the size of the area covered by the array.

The only filter parameter that is predefined is the order of the FIR filter, i.~e.~the number of filter coefficients. The filter order essentially determines the time span of the filter. Therefore, similar to the post-subtraction scheme, we found that the order can be too high, in which case the seismic array cannot provide sufficient information to disentangle NN contributions from individual seismic waves. The filter order is too low when an insufficient amount of data is used to accurately estimate the NN from individual, resolved waves. 
\begin{figure}[t]
\centering
\includegraphics[width=\columnwidth]{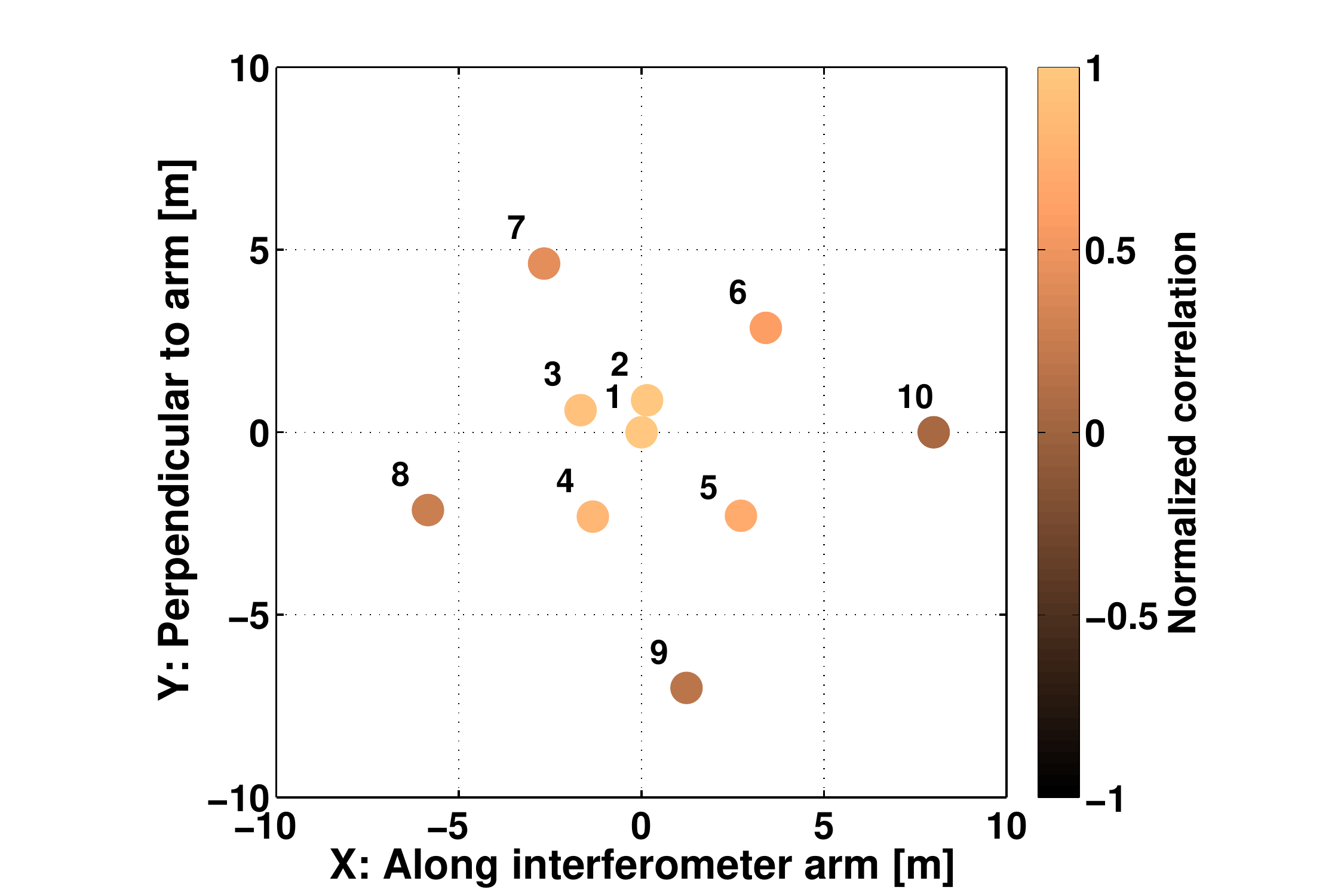}
\includegraphics[width=0.75\columnwidth]{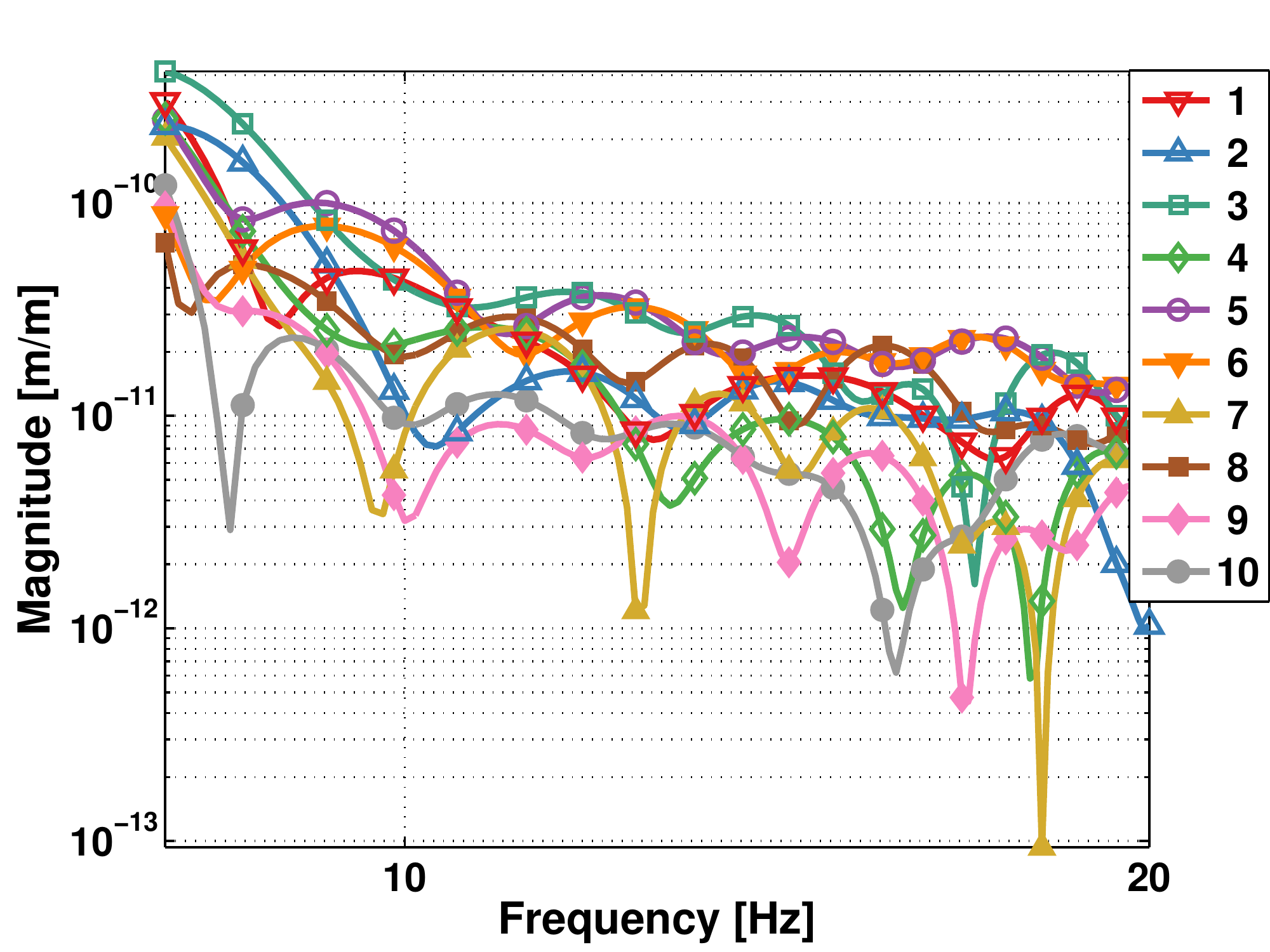}
\caption{(color online). The upper plot shows the configuration of the spiral array. The colors correspond to the normalized seismic correlation between all seismometers and seismometer 1. The numbering of seismometers corresponds to the traces in the lower plot, which shows the magnitude of the FIR filter for each sensor in units of test mass (NN) displacement over seismic displacement.}
\label{fig:bode}
\end{figure}
We will later explain why the wave nature of the seismic field still matters in the context of FF cancellation. The FIR filter that yielded sufficient subtraction in all simulation runs has order $N=50$ corresponding to a time span of 0.5\,s. The MISO FIR filter coefficients were calculated from the 100\,s long seismometer and test mass time series generated as described in Section~\ref{sec:simulation}. All time series are pre-conditioned with band-pass and whitening filters. An example of a Bode plot of the filter for a spiral array is shown in Figure~\ref{fig:bode}. The fact that for example seismometers 3 or 5 have relatively high filter magnitudes at some frequencies is interesting since their correlation with the NN is very small (as calculated by Equation~\ref{eq:CSN}). This situation can be described as a trade-off between gaining information about how NN is generated close to each seismometer (the simple local model), and gaining information about how NN integrates over the seismic field based on its wave nature. 

\begin{figure}[t]
\centering
\includegraphics[width=\columnwidth]{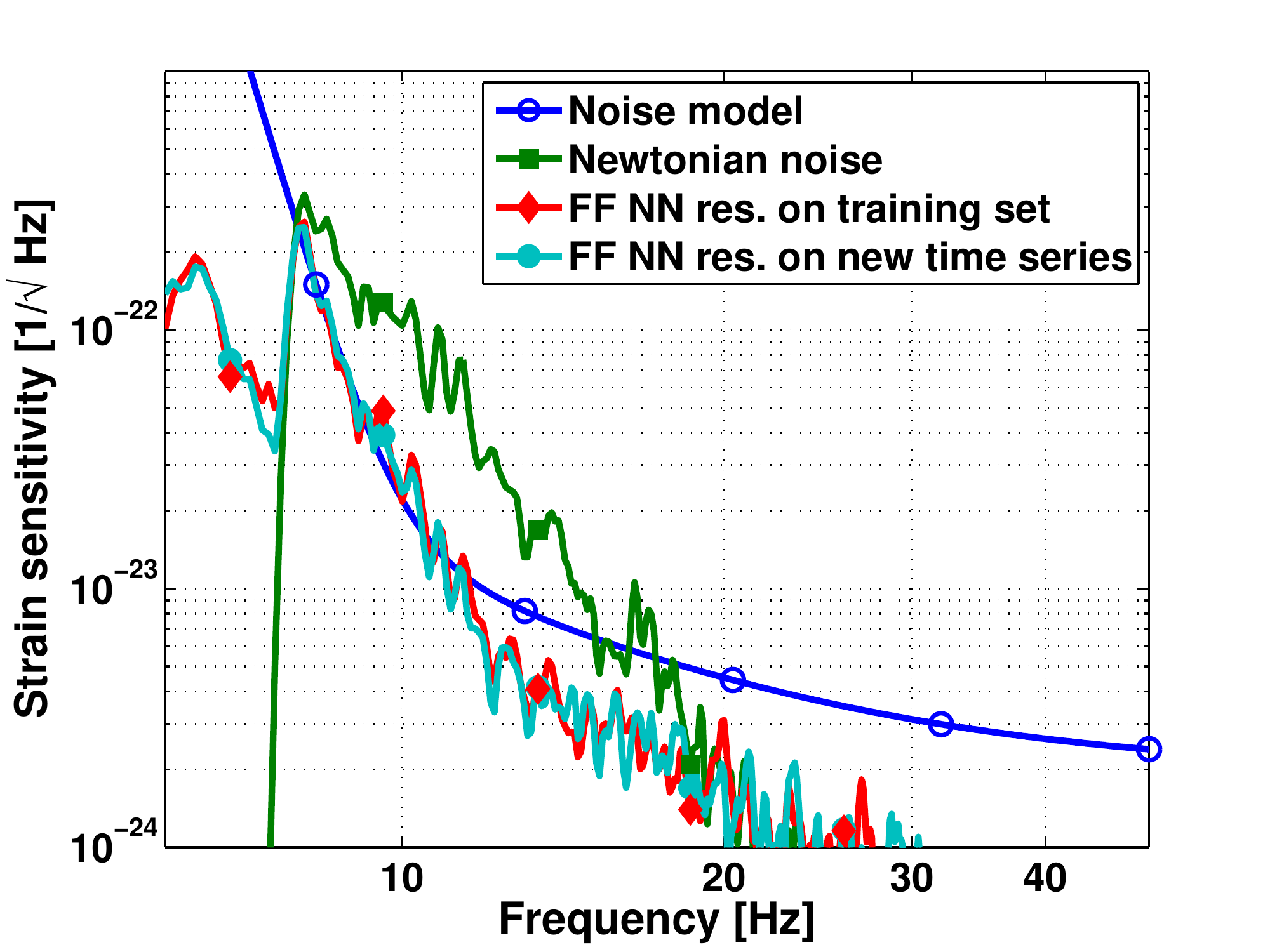}
\caption{(color online). Spectrum of simulated Newtonian noise (green line), proposed third generation sensitivity curve (blue line), and NN residuals of FF subtraction on the training set (red line), and on a second set of time series using the same filter (cyan) using the 10-sensor optimal array.}
\label{fig:preff}
\end{figure}

The FF noise cancellation performance is shown in Figure~\ref{fig:preff}. Since the FIR filter coefficients are the same for the entire time series (see Section~\ref{sec:implementation} for alternative filter implementations), we included two NN residuals, one for the Wiener filter that subtracts on the same time series used to calculate the filter coefficients, and a second one where the same filter is applied to subtract NN from another time series. 
The two time series represent different sets of local sources and wavelets. The subtraction performance is very similar for the two cases, and therefore we can conclude that subtraction performance does not depend as much on the specific wave content of the seismic field as it depends on the average correlations between sensors and the NN. While the Wiener filter applied to the data on which it was trained is an acausal use of the filter and could not be applied online, it is useful to see that the subtraction efficacy does not degrade for times that are not the training data for the filter.  As with the post-subtraction, FF cancellation performed similarly for the circular, linear and spiral arrays.

\begin{figure}[t]
\centering
\includegraphics[width=\columnwidth]{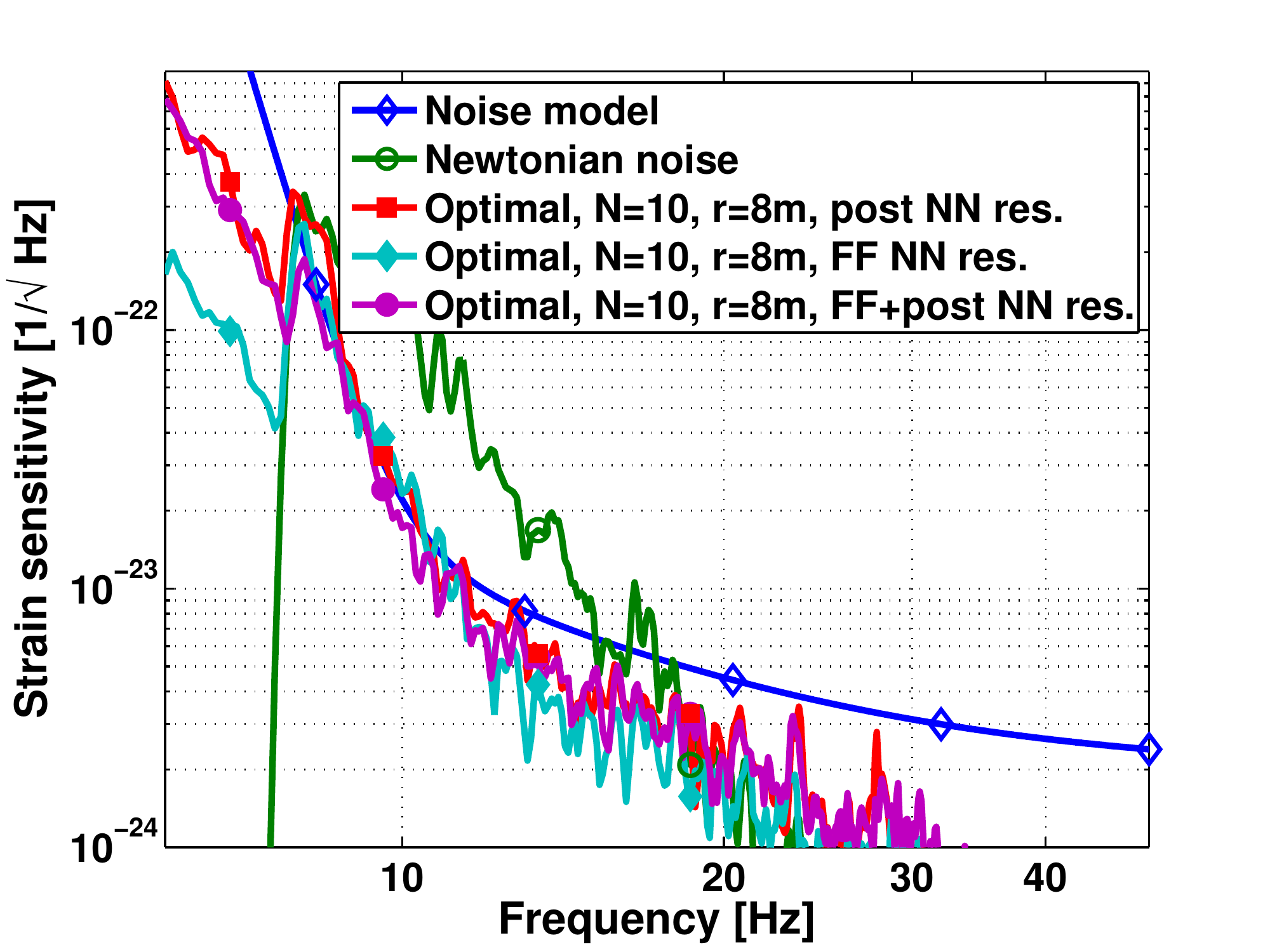}
\caption{(color online). Feed-forward Newtonian noise sub- traction efficiency for third generation LIGO detectors. Simulated NN before subtraction (green), expected strain sensi- tivity (blue line), NN residuals after subtraction using postpro- cessing (red line), online feed forward (cyan line), and both methods combined (magenta line). Note that the combination of methods is close to the same level as either method individually. This indicates that we can safely apply feed-forward subtraction in real time, and clean up any leftover noise in postprocessing if needed.}
\label{fig:postff}
\end{figure}

Finally, we investigate the possibility of combining the online FF cancellation with post-subtraction. Figure~\ref{fig:postff} shows the residual NN spectra for the three subtraction methods. 
Overall, there is no clear advantage or disadvantage to combining the two methods. When both techniques are applied, NN residuals are smaller at lower frequencies, but residuals are larger at higher frequencies. In conclusion, it was demonstrated that the standard static MISO FIR Wiener filter provides robust and sufficient subtraction results. Whereas a combination of FF and post-subtraction does not give further improvement in simulation, it could prove more effective in scenarios where strong occasional seismic disturbances leave significant residuals after FF cancellation. 

\section{Comments on the possibility of suppressing the GW signal through NN subtraction}
\label{sec:GWminusNN}

A common concern is that true gravitational wave signals could be subtracted out along with NN.  The primary reason this is not a concern is that seismic sensors are not directly sensitive to gravitational waves.  If they were, sophisticated systems such as LIGO would be unnecessary.  Here we briefly discuss several other smaller couplings between NN subtraction signals and the GW channel, and why they are not a concern.
 
\textit{1) Spurious electromagnetic coupling between test mass actuators and seismic sensors:} The LIGO detectors are controlled such that the test mass mirrors do not move in response to a gravitational wave.  Rather, the feedback forces applied to the mirrors in the GW band contain the GW information, and combinations of the individual mirror feedback forces comprise the GW channel used for LIGO analysis.  Spurious electromagnetic coupling between the actuators applying these feedback forces and the environmental monitoring sensors used for subtraction can lead to some of the GW signal being subtracted off unintentionally. This effect is very small, and can be further suppressed by measuring this coupling and correcting for it in the subtraction algorithm if necessary.

\textit{2) Ground recoil due to active seismic isolation, measured by seismic sensors:} Feedback forces actuating on the active seismic isolation structures supporting each test mass contain the GW signal at a small level.  The ground supporting the isolation system will recoil as a result of such a large mass moving.  Seismic sensors in the vicinity will measure this recoil as seismic motion, and will attempt to subtract away the NN due to this measured seismic motion.  This is a 2$^{\rm{nd}}$ order coupling effect, and the correction to the GW signal is very small, so further calculations of this effect are outside the scope of this paper.  As with the electromagnetic coupling, if necessary, this coupling can be measured and corrected for in the subtraction algorithm.

\textit{3) Earth as a GW detector:} The Earth responds to GWs with displacement amplitudes $h\frac{\beta}{\omega}$, where $h$ is the GW strain, $\beta$ is the speed of a shear wave in the ground and $\omega$ is the frequency~\cite{Dys:1969}. At all frequencies, this displacement is much smaller than is measureable by the best seismic sensors available, so this coupling is negligible.

\textit{4) Very short training times for feed-forward filters allow random correlations between transient seismic and GW events:} A feed-forward filter which has been trained on a very short data set can potentially remove signals from the original data stream.  This is because such a filter is created using correlation information between seismic sensors and the GW channel.  During a short data set, there may be random correlations between transient seismic events and true transient GW events, which would create a filter capable of subtracting away the GW event.  However, we only allow filters to be trained on data sets which are much longer than any burst or compact binary coalescence GW event that we expect, thus averaging over any seismic transients which could cause a problem for transient GW events in our data stream.  

For all of these effects, it is possible that without correction, a very small amount of GW signal could be subtracted away from the GW channel.  Once Advanced LIGO is constructed, tests can be done, such as injecting artificial GW signals into the detector, and measuring the amount by which they are suppressed by NN subtraction.


\section{Future Work}
\label{sec:implementation}
In this section, we outline the main challenges associated with NN subtraction and discuss in more detail where our numerical simulation needs to be refined to better understand the associated risks. By far the greatest challenge of NN subtraction is to make sure that all relevant sources of gravity perturbations are identified. In this context, our simulation is certainly highly simplified. However, our estimates indicate that the seismic field gives the only NN contribution that will be relevant to the advanced detectors or their potential upgrades as represented by the strain-noise model used in this paper (see Section~\ref{sec:appendix} for details). This justifies the exclusion of other NN sources in our numerical simulation. 

The more interesting aspect of the source-identification problem is whether we can be sure that all relevant degrees of freedom of the seismic field are or can be monitored. The seismic array in Section~\ref{sec:optimize} is designed based on prior knowledge. For example, in this simulation we assumed that there are no significant NN contributions from body waves that propagate through the ground in all directions. Surface waves are expected to have much larger amplitudes than body waves near 10\,Hz at the LIGO sites~\cite{HuTh1998}, but a detailed study of the fields has not yet been done. Our measured seismic spectra and NN estimates shown in Figure~\ref{fig:nnbudget} do not tell us the wave content of the 
seismic field. 
A seismic array in place at the LIGO Hanford site is currently collecting data, from 
which we should be able to determine the relative 
significance of body waves.

Another related issue that is often discussed is the scattering of seismic waves, which we assume is negligible for our simulation. This could indeed pose a severe problem to NN subtraction even if scattering is identified and fully characterized. Scattering can in principle make it impossible to estimate NN from seismic measurements at the surface since it can lead to a more complex field structure that is not completely characterized by surface displacement. Moreover, it is possible that scattered waves have higher wave numbers compared to the freely propagating surface waves, so that the density of the seismic array would need to be increased to a point where it becomes infeasible or at least very challenging to monitor the entire field accurately. However, for the modest subtraction performance that we require for the detector model we consider in this paper, the major portion of the gravitational noise perturbation comes from the surface area very close to the test mass, as opposed to higher subtraction factors, where a substantially larger surface area needs to be considered. It follows that scattering will only be a problem if it is strong enough to alter seismic waveforms significantly over very short propagation distances. Since the ground medium close to the test mass is fairly uniform, high scattering cross sections are unlikely to be observed. Seismic measurements will be necessary to test this assumption.

Methods that we have not investigated in this paper could help to mitigate some of the risks if necessary. In our simulation the FF filter used was implemented as a static Wiener filter, however it is possible to let the filter coefficients adapt slowly to changes of the seismic field. This adaptive filter technology has many applications and is well established~\cite{Say2003}. Also, once the array design has been chosen based on previous seismic measurements, cross-correlations observed with this array can help to find better array configurations. In other words, it will be possible to adapt to changing properties of the seismic field not only through adaptive filter technologies, but also through changes in the hardware configuration. 

Adding more details to the numerical simulation like scattered waves or body waves can tell us in advance how the array would have to be modified to maintain the same level of subtraction performance for these more complicated scenarios. It would be wise to investigate all scenarios irrespective of how well we think we understand the seismic fields.

We will also create and validate tests which will allow us to show that NN suppression will not adversely affect the GW detection potential of terrestrial detectors.  This may include direct artificial injections into a detector or data stream to see how the GW signal is changed with and without NN subtraction, as well as more detailed calculations and simulations of the strength of each of the couplings described in Section~\ref{sec:GWminusNN}, to confirm that none of them are a concern for 2$^{\rm nd}$ and  3$^{\rm rd}$ generation GW detectors.


\section{Conclusions}
\label{sec:conclusion}
We have shown that a relatively small number of medium sensitivity geophones or accelerometers can be used to estimate the Newtonian gravitational fluctuations 
with a reasonably high accuracy. Under our simplifying assumptions for the seismic 
fields and the structure of the ground, this allows us to use seismic data to subtract 
the gravitational noise due to seismic motion from the interferometer data stream well enough that the 
Advanced LIGO, as well as 3$^{\rm rd}$ generation detectors should not be limited by this terrestrial noise source. 

We found that the array configuration has a minor impact on the subtraction residuals. The more important design parameters are the number of seismometers, the area covered by the seismic array, and proper pre-conditioning of the time series that are used for the NN estimate. 

Our numerical simulation needs to be developed further to test subtraction of other possible contributions to the seismic field that have mostly been considered insignificant for the NN problem in advanced detectors in the past, as for example body waves and scattered waves. Testing cancellation of NN by factors of 10 or more requires a more accurate simulation of seismic fields.

The offline, acausal subtraction scheme should naturally outperform the online, adaptive causal feed-forward technique, but for the simple implementation of the post-subtraction used in this paper, the subtraction performances were comparable. To get latency for a cleaned-up data stream to be less than $\sim$1 minute, we will do initial subtraction online and then make the final subtraction offline.

These NN subtraction techniques will have a modest improvement on second generation
detectors (Advanced LIGO, Advanced Virgo, KAGRA), but the true promise will come towards
the end of the decade. At that time these techniques will be necessary to achieve the next order of magnitude improvement in astrophysical reach with third generation detectors. 


\section*{Acknowledgements}
We thank the National Science Foundation for support under grant PHY-0555406. JCD also acknowledges the 
support of an NSF Graduate Research Fellowship. We thank R.~Schofield and A.~Effler for their help with 
the measurements presented in Section~\ref{sec:appendix}. LIGO was constructed by the California Institute of  Technology 
and Massachusetts Institute of Technology with funding from the National Science Foundation and operates 
under cooperative agreement PHY-0107417. This article has LIGO internal Document Number P1200017.


%


\end{document}